\documentclass[sigconf]{acmart}

\renewcommand\footnotetextcopyrightpermission[1]{}
\acmConference[CHIWORK '26 Workshop]{%
  CHIWORK '26 Workshop: Interrogating GenAI Augmentation for CHIworkers}{%
  June 22, 2026}{Linz, Austria}
\acmYear{2026}

\newcommand{\workshopNote}{%
  \begin{quote}
    \small\itshape
    This position paper was presented at the CHIWORK '26 Workshop
    \emph{Interrogating GenAI Augmentation for CHIworkers:
    Strategies for Professional Autonomy and Accountability}
    (June 22, 2026, Linz, Austria). Workshop proposal:
    \cite{sandhaus2026interrogating}.
  \end{quote}%
}
\title{Generative AI as Support, Not Replacement, in Human-Centered Design}

\author{Amir Talakoob}
\affiliation{%
  \institution{University of North Carolina at Charlotte}
  \city{Charlotte}
  \country{USA}}
\email{atalakoo@charlotte.edu}

\renewcommand{\shortauthors}{Talakoob}

\begin{abstract}
Traditional human-centered design is grounded in human needs, behaviors, and experiences, especially throughout the development of products and services. Recent advances in artificial intelligence and machine learning, accelerated by the public launch of ChatGPT in November 2022, have created a wave of AI adoption across research, industry, and design practice. While some of this rapid adoption reflects the current enthusiasm and overuse surrounding AI tools, it has also introduced lasting changes to how designers and researchers understand users, generate ideas, evaluate systems, and make design decisions. This position paper examines how AI is reshaping human-centered design and argues that some of these changes are likely to extend beyond short-term hype. It also considers the risks and limitations introduced by this shift, including over-reliance on AI, reduced human judgment, bias, data security concerns, and unclear accountability. Finally, the paper suggests that AI literacy, careful ethical judgment, and clearer boundaries for AI use are necessary for preserving the human-centered focus of design.
\end{abstract}


\begin{document}
\maketitle

\workshopNote

\section{Introduction}

This position paper focuses on Generative AI in traditional human-centered design and discusses how the recent growth of AI is affecting design research, practice, and education. Artificial Intelligence and Machine Learning have existed for a long time, but the public release and rapid adoption of Generative AI tools created a different kind of change. Unlike many traditional AI or ML systems that are mostly used for prediction, classification, or automation, Generative AI can directly take part in tasks such as brainstorming, writing, summarizing, prototyping, and interpreting information. Because of this, GenAI does not only affect the final product or system; it can also affect the design process itself.

The current use of Generative AI has created both positive and negative opinions. On the positive side, these tools can help designers and researchers explore ideas faster, organize information, generate early prototypes, and support communication. However, the negative concerns are also important, especially in a field such as human-centered design. Overusing AI can reduce direct engagement with users, weaken human judgment, create issues of bias and accountability, and make the design process less connected to real human experiences.

A central factor in human-centered design is the focus on human needs, behaviors, experiences, and interactions. For this reason, the increasing use of Generative AI creates an important question: how can HCD practitioners benefit from AI tools without losing the human-centered values that define the field? This paper argues that Generative AI should not replace traditional human-centered design practices, but should be used as a supporting tool. The goal should not be to blindly accept AI or completely reject it, but to use it carefully in a way that keeps human experience, user evidence, ethical responsibility, and human judgment at the center of the design process.

\section{Background and Motivation}
\subsection{Traditional HCD: Strengths and Limitations}
Traditional Human-Centered Design remains valuable because it forces designers to begin with people rather than with technology. In a course-based HCD project on DoorDash transparency cards, this was especially important because the problem was not simply that DoorDash uses algorithms, but that users experience those algorithmic decisions through moments of confusion, stress, and uncertainty. The ordering user does not directly see the logic behind fee calculations, delivery estimates, restaurant rankings, item markups, or other platform decisions. Instead, they encounter these systems through interface elements such as the subtotal, delivery fee, estimated tax, tip, and final total. Traditional HCD gave our team a structured way to move from this broad concern about algorithmic opacity to more specific user needs by using interviews, contextual inquiry, affinity diagrams, flow diagrams, sequence diagrams, personas, sketches, storyboards, prototypes, and user evaluation.

One of the main strengths of traditional HCD is that it reveals context-of-use. In the early milestones, our team did not start by designing the transparency card immediately. Instead, we first identified relevant stakeholders, including ordering users and Dashers, and then used interviews and contextual inquiry to understand where confusion appeared in the actual ordering process. This was important because users did not experience DoorDash in an abstract or neutral setting. They often used it when they were hungry, busy, under time constraints, working, studying, or trying to make a quick decision. These details mattered because they changed what kind of transparency would be useful. A long technical explanation of algorithmic pricing might be ``transparent'' in one sense, but it would not be human-centered if it slowed down a user who only has a short lunch break. This showed one major strength of traditional HCD: it connects design decisions to real situations rather than assuming that more information automatically means a better experience.

Traditional HCD also helped us turn scattered user observations into design requirements. Through the affinity diagram and the consolidated flow and sequence diagrams, our team moved from individual interview notes to broader categories such as pricing and fees, wait times, decision-making, trust, transparency, and system clarity. This process made the design problem more manageable. Instead of treating ``DoorDash transparency'' as one large issue, we were able to identify where transparency was most needed: fee breakdowns, item markups, wait-time explanations, and moments where the user needed to make a decision. This was one of the clearest strengths of the HCD process in our project. It created a bridge between qualitative user experiences and concrete design requirements, such as allowing users to understand how prices are calculated, distinguish between menu prices and markups, interpret delivery updates, and make informed choices without unnecessary confusion.

Another strength of traditional HCD is its iterative nature. Our project did not move directly from user interviews to a final interface. Instead, the design changed across milestones. In the persona, sketch, and storyboard phase, we created and critiqued multiple design directions before consolidating them into a final persona and design concept. The final persona represented a busy, budget-conscious user with limited time, which pushed the design toward concise and quickly accessible transparency cards rather than dense explanations. This step was important because it made the design emotionally and practically grounded. The transparency card was not only a technical explanation of fees; it was a response to a user who wants to know why the final price is higher, whether the order is still worth it, and whether they should continue.

The later evaluation milestones showed another strength of HCD: it can expose design flaws before a system is fully built. Through think-aloud testing and heuristic evaluation, we found that the transparency card improved understanding but did not fully solve the user’s problem. The low-fidelity prototype helped explain why the final cost was higher, but the heuristic evaluation showed that this explanation did not provide enough user control. If a user learns that the order is too expensive, the system still needs to support meaningful next steps, such as editing the cart, switching to pickup, choosing another restaurant, or adjusting the tip. This was a critical finding because it showed that transparency alone is not enough. A design can explain the system and still leave the user frustrated if the explanation leads to a dead end. In this sense, traditional HCD helped us move from ``users need information'' to a more mature requirement: users need actionable transparency.

However, our project also showed several limitations of traditional HCD. The first limitation is that traditional HCD can be slow and resource-intensive. Each milestone required a different method: interviews, contextual inquiry, diagramming, persona creation, sketching, storyboarding, low-fidelity prototyping, think-aloud testing, heuristic evaluation, high-fidelity prototyping, and a quantitative evaluation plan. Each step added value, but each step also required time, coordination, participant access, and interpretation. Within the scope of a course-based project, this was manageable but still difficult. In a real product environment, especially one involving fast-moving algorithmic systems, this process may struggle to keep up with the pace of platform changes.

A second limitation is that traditional HCD depends heavily on the quality and availability of participants. Our sample was limited, and many potential participants either did not use DoorDash frequently or had different levels of familiarity with the platform. This made it harder to generalize findings. It also created a tension between designing for the users we could access and designing for the broader population of DoorDash users. Traditional HCD values situated user feedback, but if the participant pool is narrow, the resulting requirements may reflect only a partial view of the user experience. This became especially clear when our team initially considered both ordering users and Dashers but later narrowed the project mainly to ordering users. That decision helped the project become more focused, but it also left part of the algorithmic system less explored.

A third limitation is that traditional HCD can struggle with opaque algorithmic systems because users and designers may only see the interface layer, not the full decision-making logic behind it. In the DoorDash case, users can observe that prices, fees, rankings, discounts, and delivery estimates change, but they cannot easily know why. Designers can interview users about their confusion, but they may not have access to the proprietary model or business logic that produces that confusion. This means traditional HCD can identify where opacity harms user trust, but it cannot always explain the system’s internal behavior. As a result, the design intervention may become a transparency approximation rather than full transparency.

Overall, the project showed that traditional HCD is strongest when it grounds design in real user context, translates qualitative observations into requirements, and tests designs before full implementation. At the same time, it is limited by time, participant access, sample size, and the difficulty of studying black-box algorithmic systems from the outside. In the DoorDash project, traditional HCD helped us understand that the problem was not only hidden information, but hidden information in a high-stress decision-making context. Its main lesson was that good transparency must be concise, contextual, and actionable. However, its main limitation was that reaching this conclusion required a long process, and even then, the design could only respond to the algorithmic system from the user-facing side.

\subsection{Historical Parallels: GenAI, CGI, and Telephone Lines}
At this point in time, AI systems and machine learning models are moving from a mostly theoretical or specialized field into a more practical and everyday usable stage. This shift has mainly happened because of recent improvements in computing power, storage, and access to AI tools by both research labs and the general public. However, there are still many conflicting opinions with regard to GenAI. A lot of these concerns come from data privacy, intellectual property, and questions about AI-generated content, whether in the form of art, text, or code. In academic settings, undisclosed AI-generated text has also caused serious concerns about plagiarism and cheating. This has created an arms race atmosphere, where AI-detection tools are trying to keep up with newer models, paraphrasing tools, and auto-typers.
We should also consider the fact that these generated contents are still in the early stages of development and, similar to other technologies in the past, they are prone to public backlash during early adoption. This does not mean that the backlash is meaningless. In many cases, public resistance shows real concerns about control, trust, authenticity, ownership, and how much space a new technology should occupy in daily life. Later in this paper, I discuss AI literacy and proper guidelines, because without them, opposition to the uncontrolled use of AI becomes the only possible response.

One useful historical parallel is Computer Generated Imagery (CGI) in the movie industry. Before the first major use of CGI in \textit{Tron} in the 1980s, movies tended to use practical effects, stop-motion animation, matte paintings, and miniatures. It was really interesting for me to learn that ``The film was even disqualified from the best special effects Oscar, since using computers was considered 'cheating'.'' \cite{cgi} But now, CGI is an inseparable part of the movie industry. This example is useful because it shows how a technology that was once seen as artificial or even unfair later became a normal part of creative production. At the same time, this does not mean every use of CGI is good. The lesson is not that all new technologies should be accepted without criticism, but that their long-term value depends on how they are used, regulated, and understood.

A second example is the public reaction to utility and telephone poles during their early adoption. A while ago, I came across the following piece of history regarding the early days of telephone poles in New York: ``By 1889, The New York Times was reporting a 'War on Telephone Poles.' Wherever telephone companies were erecting poles, homeowners and business owners were sawing them down, or defending their sidewalks with rifles.'' \cite{biss2009telephonepoles}
It is also important to consider the early design and condition of utility poles, as they ``carried a wire for each telephone-sometimes hundreds of wires. And in some places there were also telegraph wires, power lines, and trolley cables. The sky was filled with wires.'' \cite{biss2009telephonepoles}

This example connects strongly to the current atmosphere around GenAI. Telephone poles represented progress, communication, and connection, but they also created visible disruption in public space. Similarly, GenAI creates new possibilities for writing, design, research, and communication, but it also creates disruption in academic work, creative labor, authorship, and trust. The point of this comparison is not to claim that AI is the same as telephone lines or CGI. Instead, it shows that early technological adoption often includes both real benefits and real concerns. For HCD, this is important because the goal should not be blind adoption or complete rejection. The goal should be to understand how GenAI changes human experience and then create responsible practices around that change.
\begin{figure}
    \centering
    \includegraphics[width=0.5\linewidth]{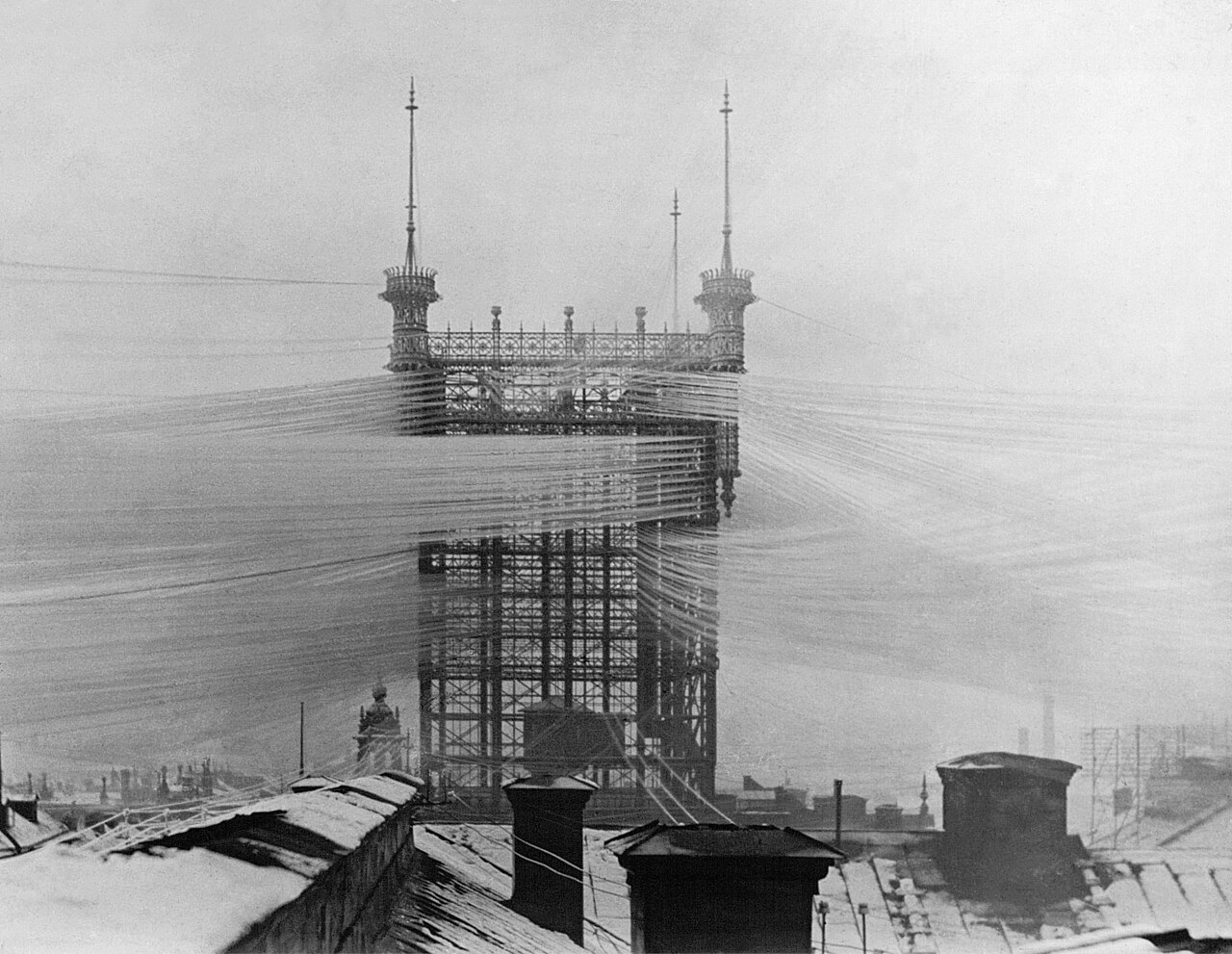}
    \caption{The ``Telephone Tower'' with hoarfrost built by Stockholms Allmänna Telefon AB at Malmskillnadsgatan 30 in Stockholm during the 1890s \cite{tekniskamuseet1890telephone}.}
    \label{fig:telephone-tower}
\end{figure}

\section{Position and Argument}
\subsection{Uses of GenAI in HCD}
Earlier in this paper, I discussed how our team used traditional HCD in a course-based HCD project, but during this process, and through the role of GenAI advocate, our group also explored several possible uses of GenAI. These included persona creation, using GenAI as an alternative heuristic evaluator, and, in some cases, using it as a writing quality checker. These examples show that GenAI can enter the HCD process at different stages, not only as part of the final product but also as part of research, design, evaluation, and communication.

One paper that I wish I had encountered during our project’s literature review was ``From Generation to Simulation: Responsible Use of AI Personas in Human-Centered Design and Research'' \cite{10.1145/3772363.3778745}. This is a strong example of how closely our course-based project reflected real-world struggles in HCD. The workshop directly focuses on the responsible use of AI-generated and AI-simulated personas in human-centered design and research, which connects closely to our own use of personas and our later concerns about whether AI-generated user representations can be useful without replacing real user evidence.

During my earlier experience presenting at the International Conference of Computing in Civil Engineering (I3CE) as a computer science student, I was surprised by the number of workshops and papers focusing on AI and AI literacy for Civil Engineering and Construction Management students. This past experience gave me the idea to check the ACM CHI 2026 conference program to better understand the changes in the HCI field and the recent uses of AI, both generative and in general.

A brief review of the ACM CHI 2026 program shows that AI was not a marginal topic; even the first day included dozens of AI-focused papers, sessions, workshops, panels, and meetups across areas such as AI governance and accountability, AI in work and expertise, AI-driven creativity and design methods, human-LLM collaboration, AI personas, vibe coding, AI literacy, and standards for LLM use as simulated research participants \cite{chi2026}.

This broad presence reflects the current AI wave in the research community, including both genuine long-term interest and some degree of hype. In my opinion, some of this attention will deflate, but AI as a tool will remain for a long time, just as computers and the internet did.

For HCD, this means that the question is no longer whether GenAI will enter the field. It already has. The more important question is how it should be used. GenAI can support activities such as persona creation, ideation, prototyping, evaluation, writing, and sensemaking, but these uses also need to be treated carefully. If AI-generated personas, evaluations, or summaries are accepted without checking them against real users and real contexts, then HCD can become less human-centered while appearing more efficient. Therefore, the value of GenAI in HCD depends on whether it is used as a support for human-centered work, rather than as a replacement for human evidence and designer judgment.

\section{Implications and Discussion Points}

\subsection{Data Security}
There are many concerns regarding the use of GenAI, but one that directly affects researchers and HCD practitioners is the use of AI and LLMs on sensitive and unpublished data. In ``All Accept, No Reject: Evaluating LLMs as `Peer' Reviewers'' \cite{10.1145/3772318.3791300}, this concern is mostly discussed through the lens of review quality and whether LLMs can provide reliable peer-review feedback. However, I also look at this issue from the perspective that sensitive information may be exposed to commercial AI models that collect or process user data without the consent of the authors. This is especially important in academic settings, where unpublished papers, research ideas, datasets, student work, and confidential review materials may be entered into AI tools for summarization, editing, or evaluation.

This creates a serious problem for both data security and intellectual ownership. If a researcher uploads an unpublished manuscript or a reviewer uploads a confidential paper to an external AI system, the issue is not only whether the AI gives good feedback. The issue is also whether the authors agreed for their work to be processed by that system in the first place. In this sense, GenAI introduces a new kind of risk into academic and design work: private or unfinished ideas can be processed by external commercial systems before they are published, protected, or properly credited.

For HCD, this is especially concerning because user interviews, survey responses, usability notes, and project materials often contain personal or contextual information that should not be treated as generic input for an AI system.

\subsection{AI Literacy}
AI literacy should be considered a vital part of today's academic and research environment. Students and researchers must understand the basic nature of Generative AI models, including their limitations and the distinction between correlation and causation. This does not mean that every researcher needs to become an AI expert, but they should understand that GenAI is not a genie or an objective reasoning system. It generates responses based on patterns in data and the prediction of plausible next tokens.

One useful way to build this understanding could be through learning how to train a miniature GPT-like model from scratch \cite{book}. Even a small-scale version would help users understand why these models can produce confident answers without actually verifying truth, observing the world, or performing real-time tasks. For example, if a user asks an AI model to ``time me when I run a mile,'' the model may respond with a typical answer such as ten minutes, even though it is not actually measuring time. This shows how easily users can misunderstand what GenAI is capable of doing.

For HCD, this is especially important because designers may use GenAI to summarize user feedback, create personas, evaluate interfaces, or generate design alternatives. Without AI literacy, these outputs may be confused with real user evidence. Therefore, AI literacy should include not only how to use AI, but also when to verify it, when not to use it, and when human judgment and direct user research must remain central.

\subsection{GenAI as Support, Not Replacement}
Overall, Generative AI should not be viewed as a replacement for traditional human-centered design. HCD remains valuable because it keeps design grounded in real human needs, behaviors, experiences, and interactions. However, GenAI can still support this process when it is used carefully. It can help with brainstorming, organizing information, summarizing feedback, creating early design alternatives, and improving communication.

The main risk is that designers may begin to treat AI-generated outputs as a substitute for real user evidence or human judgment. This would weaken the human-centered nature of the design process. For this reason, the future of HCD should not be based on blind acceptance or complete rejection of GenAI. Instead, practitioners should use GenAI as a supporting tool, while keeping user research, ethical responsibility, AI literacy, and human judgment at the center of design work.

\begin{acks}
  \textbf{AI Use Statement:} Generative AI was used as a supporting tool during the writing process of this paper. Its main role was to help with organizing ideas, improving sentence clarity, and checking whether the argument was consistent across sections. However, the main reflection, examples, project experience, and final decisions about what to include remained my own. I did not use GenAI as a replacement for course materials, project experience, or personal analysis. Instead, it was used in the same way that this paper argues GenAI should be used in HCD: as a support tool that can assist the process without replacing human judgment, responsibility, or direct experience.

\end{acks}

\bibliographystyle{ACM-Reference-Format}
\bibliography{references}

@inproceedings{sandhaus2026interrogating,
  author    = {Sandhaus, Hauke and Imteyaz, Kashif and Almutairi, Mohammed
               and Prajod, Pooja and Ramesh, Divya and Savage, Saiph
               and Yang, Qian and Muller, Michael},
  title     = {Interrogating {GenAI} Augmentation for {CHIworkers}:
               Strategies for Professional Autonomy and Accountability},
  year      = {2026},
  isbn      = {979-8-4007-2598-2/2026/06},
  publisher = {Association for Computing Machinery},
  address   = {New York, NY, USA},
  url       = {https://doi.org/10.1145/3805029.3818271},
  doi       = {10.1145/3805029.3818271},
  booktitle = {Adjunct Proceedings of the 5th Annual Symposium on
               Human-Computer Interaction for Work},
  series    = {CHIWORK '26},
  location  = {Linz, Austria},
  month     = jun,
}

@misc{cgi,
  author = {Rose, Steve},
  title = {{`Frankly it blew my mind': How Tron changed cinema -- and predicted the future of tech}},
  year = {2022},
  month = jul,
  day = {5},
  publisher = {The Guardian},
  url = {https://www.theguardian.com/film/2022/jul/05/tron-steven-lisberger-interview},
}

@misc{biss2009telephonepoles,
  author = {Biss, Eula},
  title = {{The War on Telephone Poles}},
  year = {2009},
  month = feb,
  publisher = {Harper's Magazine},
  url = {https://harpers.org/archive/2009/02/the-war-on-telephone-poles/},
  note = {Accessed: 2026-05-15}
}

@misc{tekniskamuseet1890telephone,
  author = {{Tekniska museet}},
  title = {{Telephone Tower with hoarfrost in Stockholm}},
  year = {1890},
  publisher = {Wikimedia Commons},
  url = {https://commons.wikimedia.org/wiki/File:Telephone_Tower_with_hoarfrost_in_Stockholm.jpg},
  note = {Image. Unknown photographer. Original kept in the National Museum of Science and Technology, Sweden.}
}

@inproceedings{10.1145/3772363.3778745,
    author = {Kocaballi, A. Baki and Prpa, Mirjana and Salminen, Joni and Amin, Danial and J Jansen, Bernard},
    title = {From Generation to Simulation: Responsible Use of AI Personas in Human-Centered Design and Research},
    year = {2026},
    isbn = {9798400722813},
    publisher = {Association for Computing Machinery},
    address = {New York, NY, USA},
    url = {https://doi.org/10.1145/3772363.3778745},
    doi = {10.1145/3772363.3778745},
    booktitle = {Proceedings of the Extended Abstracts of the 2026 CHI Conference on Human Factors in Computing Systems},
    articleno = {945},
    numpages = {5},
    location = {
    },
    series = {CHI EA '26}
}

@misc{chi2026,
  author = {{ACM SIGCHI}},
  title = {{CHI '26 Program: CHI Conference on Human Factors in Computing Systems}},
  year = {2026},
  publisher = {Conference Programs},
  url = {https://programs.sigchi.org/chi/2026/program/all?itemsType=SESSION&sortDirection=asc&sortType=TIME&viewType=LIST}
}

@inproceedings{10.1145/3772318.3791300,
    author = {Verma, Nitin and Landrum, Asheley R.},
    title = {All Accept, No Reject: Evaluating LLMs as “Peer” Reviewers},
    year = {2026},
    isbn = {9798400722783},
    publisher = {Association for Computing Machinery},
    address = {New York, NY, USA},
    url = {https://doi.org/10.1145/3772318.3791300},
    doi = {10.1145/3772318.3791300},
    booktitle = {Proceedings of the 2026 CHI Conference on Human Factors in Computing Systems},
    articleno = {1573},
    numpages = {14},
    location = {
    },
    series = {CHI '26}
}

@incollection{book,
  author = {Chollet, Fran{\c{c}}ois and Watson, Matthew},
  title = {{Text generation}},
  booktitle = {{Deep Learning with Python, Third Edition}},
  publisher = {Manning},
  year = {2025},
  url = {https://deeplearningwithpython.io/chapters/chapter16_text-generation/}
}

\end{document}